\documentclass[aps,prd,showpacs,twocolumn,groupedaddress]{revtex4}

\usepackage{graphicx}   
\usepackage{amsmath,amssymb,bm}

\newcommand{\lvk}{\langle0_\mb{k}|}
\newcommand{\rvk}{|0_\mb{k}\rangle}

\newcommand{\Oh}{\hat{O}}
\newcommand{\Ph}{\hat{P}}
\newcommand{\Hh}{\hat{H}}
\newcommand{\mb}{\mathbf}

\newcommand{\mr}{\mathrm}

\newcommand{\su}[1]{\sideset{}{_{#1}}\sum}
\newcommand{\ok}[1]{\omega_{\mb{k}#1}}
\newcommand{\fk}[1]{f_{\mb{k}#1}}
\newcommand{\fbk}[1]{\overline{f}_{\mb{k}#1}}
\newcommand{\Ahk}[1]{\hat{A}_{\mb{k}#1}}
\newcommand{\Ahdk}[1]{\hat{A}^{\dag}_{\mb{k}#1}}

\def\p {\partial}

\def\be {\begin{equation}}
\def\ee  {\end{equation}}
\def\bea {\begin{eqnarray}}
\def\eea {\end{eqnarray}}
\def\nn {\nonumber}

\begin{document}

\title{Ultraviolet behavior in background independent quantum field theory}

\author{Viqar Husain}
\email[]{vhusain@unb.ca}

\author{Andreas Kreienbuehl}
\email[]{k7cw3@unb.ca}

\affiliation{University of New Brunswick\\
Department of Mathematics and Statistics\\
Fredericton, NB E3B 5A3, Canada}

\date{\today}

\begin{abstract}

We describe  a background independent quantization of the scalar field that
provides an explicit  realization of Fock-like states and  associated operators in a 
polymer Hilbert space. The vacuum expectation values of the commutator and anti-commutator of the creation and annihilation operators become energy dependent, and  exhibit a surprising  transition to fermionic  behavior at high  energy.  Furthermore the approach yields a modified dispersion relation with a leading correction proportional to the momentum cubed.  These results suggests a fundamental change in the ultraviolet properties of quantum fields.

\end{abstract}

\pacs{ }

\maketitle

\section{Introduction}
 
Questions concerning the behavior of spacetime   and of quantum fields  at short distances are intimately connected \cite{ISH-strucQG}. In the presence of  gravity,  usual quantum field theory (QFT) on a fixed background cannot be  a complete theory because the effect on spacetime curvature of  sufficiently high energy particles cannot be ignored. This problem is already manifest if we consider the gravitational  effect of the  QFT vacuum. This vacuum is the sum
of  ground state oscillator energies, one for each point in momentum space, and so adds up
to quite a bit, namely, the cosmological constant problem.

In addition to this ``static'' problem,  the QFT vacuum is considered to be filled with virtual particle-antiparticle pairs. The energy  density of such a pair  is its rest mass $2m$ divided by the Compton volume 
\begin{equation*} 
  \rho_{\rm pair}^{\rm rest} =  \frac{2m}{[\hbar/(mc)]^3} \simeq m^4.
\end{equation*}
If the virtual pair interacts gravitationally,  applying Newton's law gives an enhanced interaction energy density 
\begin{equation*}
\rho_{\rm pair}^{\rm int}=  \frac{Gm^2}{\hbar/(mc)}\ \frac{1}{[\hbar/(mc)]^3} \simeq m^6.
\end{equation*}
These energy densities induce virtual curvature  of spacetime via the Einstein equations.  
 Thus, it is apparent from simple arguments that high energy probes would drastically affect spacetime and make the semiclassical approximation inapplicable. 
  
There have been numerous attempts to modifying the high energy behavior of  quantum fields to 
overcome the problem sketched above. All of them have the feature that an ultraviolet cut-off is introduced
by some argument that leads to  well behaved propagation of quantum fields at high energies. In addition, all attempts involve the introduction of a fundamental length motivated by quantum gravity, whether it is string theory, loop quantum gravity (LQG) \cite{lqg-AL, lqg-T, lqg-S,lqg-R}, non-commutative geometry \cite{DFR} or some other
approach independent of these. 

In this paper we revisit the question about the high energy behavior of quantum fields. We explore not
so much a specific theory but a quantization procedure that, unlike the Schr\"odinger quantization, comes with a  specified mass scale. This is the so-called polymer or background independent quantization  method that originated in LQG. Its central feature for our purposes is that {\it the Hilbert space used for the quantization  has an inner product, which is independent of the background metric}, hence the name.  

This property does not, however, imply  that the expectation values of all operators are metric independent;  there could still be a dependence on the metric through the operators themselves (such as the Hamiltonian), or possibly through a special quantum state in which metric dependence arises through coefficients in a linear combination. We consider here  expectation values of operators in a suitably defined vacuum state. We will see that they  turn out to be metric independent. Furthermore, the mode  frequency  and the mass scale associated with the quantization give rise to a dimensionless parameter. Expectation values can then be expressed in terms of this parameter, thereby allowing us to probe their high energy behavior.  

A similar approach to the one we employ here has been applied to  black holes and cosmology \cite{HW-cosm, 
HW-qbh1, HW-qbh2, HHS-cosm}, and it has been used to derive an effective scalar field theory on a Minkowski background \cite{HHS-sft} using semiclassical states. The present work may be viewed as  a parallel but independent development.  

The outline of this paper is as follows. In Sec. II we give the quantization procedure, together with a 
definition of Fock states as well as creation and annihilation operators. This construction is new and the way it fits in the context of earlier work on the relation between Fock and polymer quantization is discussed. In Sec. III we give a computation of the expectation values of the commutator and anti-commutator for our Fock-like states. Unlike in usual QFT, these acquire a frequency dependence. The main result of his section is that  the  expectation values behave like the usual Fock states at low energies but show a surprising fermionic nature  at high energies.  Section IV contains a discussion of the results  and some open directions and speculations. 

 \section{Classical theory and polymer variables}

Although our primary interest is the free massless scalar field theory on the
Minkowski  spacetime,  let us first consider the more general metric 
\begin{equation*}
g_{\alpha\beta} \mathrm{d}x^{\alpha}\mathrm{d}x^{\beta} = -\mathrm{d}t^2 + q_{ab}\mathrm{d}x^a \mathrm{d}x^b , 
\end{equation*}
where  $q_{ab}$ denotes
the spatial metric. The canonical phase space  variables are
$(\phi,p)$ and the scalar matter Hamiltonian is given by
\begin{equation*} 
H = \int_V\frac{\sqrt{q}}{2} \left( \frac{p^2}{q} +
q^{ab}\p_a \phi\p_b\phi  + m^2\phi^2 \right)\mathrm{d}^3x, 
\end{equation*}
where $ q = \det(q_{ab})$. The  usual (Klein-Gordon) wave equation for the scalar field
follows from this via Hamilton's equations of motion.

Let us consider the phase space functions
\begin{equation}\label{basic-obs}
\begin{split}
\Phi_f (t,\mb{x}) &= \frac{1}{V} \int_V  \sqrt{q} f(\mathbf{x}-\mathbf{x}')  \phi(t,\mathbf{x}')\ \mathrm{d}^3x', \\
U(t,\mathbf{x}) &= \mathrm{e}^{i\lambda p(t,\mathbf{x})/\sqrt{q}} 
\end{split}
\end{equation}
as the new set of  variables that are to be realized as
the basic operators in the quantum theory. The function  $f$ is scalar valued, 
$\lambda$ is a  real constant with the dimension of  length
squared (in natural units), and $V$ is the volume of $3-$space.
The factors of $\sqrt{q}$ are necessary to balance density weights in
the integral and in the exponent. The variables in Eq. \eqref{basic-obs} satisfy the (equal time)
canonical Poisson bracket
\be 
\{ \Phi_f(t,\mathbf{x}), U(t,\mathbf{x}') \}  = i \frac{\lambda}{V} f(\mathbf{x}-\mathbf{x}')
U(t,\mathbf{x}'). \label{basicpb} 
\ee
These variables may be viewed as  ``dual"  to
those used  in the polymer quantization of a particle system
\cite{AFW} motivated by LQG,
where  the exponentiated configuration variable is used as the new
variable. The dual for quantum mechanical systems has been discussed
in  \cite{halvor}.  We note, however, that there are important
differences with the variables used for the polymer scalar field quantization
discussed in  \cite{QSD4,MV-fock, ALS-fock,Sahl-sft},  which do not use the available
background metric in their definition. The functions we employ are in a sense more conventional
in that they are just the field configuration and field translation variables. 
As we will see, the essential difference to the standard quantization lies in the
realization of  the field momentum operator. 

A localized field  may be defined by taking, for example, $f$ to be
a Gaussian, which is sharply peaked at a point $\mathbf{x}_j$. This means that
 \begin{equation}\label{gauss} 
 f(\mathbf{x}_j-\mathbf{x}')=\mathrm{e}^{-(\mathbf{x}_j-\mathbf{x}')^2/\sigma^2} , 
 \end{equation}
 where $\sigma^2\ll 1$. We   will assume
this in the following. 

Since our main interest in this paper is in QFT on flat spacetime, we use the Euclidean metric $e_{ab}$ and set $q_{ab} = e_{ab}$ from this point on. The generalization to arbitrary metrics is
straightforward and is of potential use for cosmology and black hole
physics.  

\section{Quantization}

The representation of the functions defined in Eq. \eqref{basic-obs} as operators on a
suitable Hilbert space is  related to that used in polymer quantum mechanics \cite{AFW,
halvor}.   We therefore briefly review this quantization before generalizing
it to field theory.  As already mentioned, a central feature of this quantization
is that it introduces a length scale in addition to $\hbar$.

The Hilbert space  is the space of almost
periodic functions, where a wave function is written as the linear
combination
\begin{equation*} 
\psi(\mb{p}) = \sum_{j=1}^{N} c_j \mathrm{e}^{i\mb{p}\mb{x}_j} = \sum_{j=1}^{N} c_j
\langle \mb{p}|\mb{x}_j\rangle. 
\end{equation*}
Here, the set of points $\{\mb{x}_j|j=1,\ldots,N\}$ is a subset (graph) of
$\mathbb{R}^3$.  The inner product is
\begin{align*}
\langle \mb{x}_j|\mb{x}_l\rangle &= \lim_{T\to\infty} \frac{1}{(2T)^3}\int_{-T}^T\int_{-T}^T\int_{-T}^T
\mathrm{e}^{-i\mb{p}(\mb{x}_j-\mb{x}_l)}\ \mathrm{d}^3p\\
&= \delta_{\mb{x}_j,\mb{x}_l}, 
\end{align*}
in which plane waves are normalizable (the right hand side is the
generalization of the Kronecker delta to continuous indices). The configuration and translation 
operators $\hat{\mb{x}}=-i\nabla_{\mb{p}}$ and $\hat{u}=\widehat{\mathrm{e}^{i\boldsymbol{\lambda}\mb{p}}}$ act
as
\begin{equation*}
\hat{\mb{x}}\mathrm{e}^{i \mb{p}\mb{x}_j} = \mb{x}_j \mathrm{e}^{i \mb{p}\mb{x}_j}, \ \ \ \ \ \
\hat{u}\mathrm{e}^{i\mb{p}\mb{x}_j}= \mathrm{e}^{i\mb{p}(\mb{x}_j -\boldsymbol{\lambda})}.
\end{equation*}
However, the usual momentum operator $\hat{\mathbf{p}}=-i\nabla_{\mb{x}}$ is not defined.
Only \emph{finite} translations can be realized and the \emph{infinitesimal} translations do not exist
in this Hilbert space. 

The polymer quantum mechanics introduced just now has been applied to the harmonic
oscillator \cite{AFW}, to the potentials $1/|\mb{x}|$ \cite{HLW-hyd} and $1/|\mb{x}|^2$ \cite{KLZ}, as well as to the cosmology and black hole cases mentioned earlier.
What is missing, however, is a complete application to QFT that goes beyond
the effective approach in Ref. \cite{HHS-sft} and that would allow for a computation of the
vacuum expectation values of field operators. This is the program we initiate
in this paper.

\subsection{Field theory}

\begin{figure} 
  \vspace{0.8cm}
  \begin{center}
\includegraphics {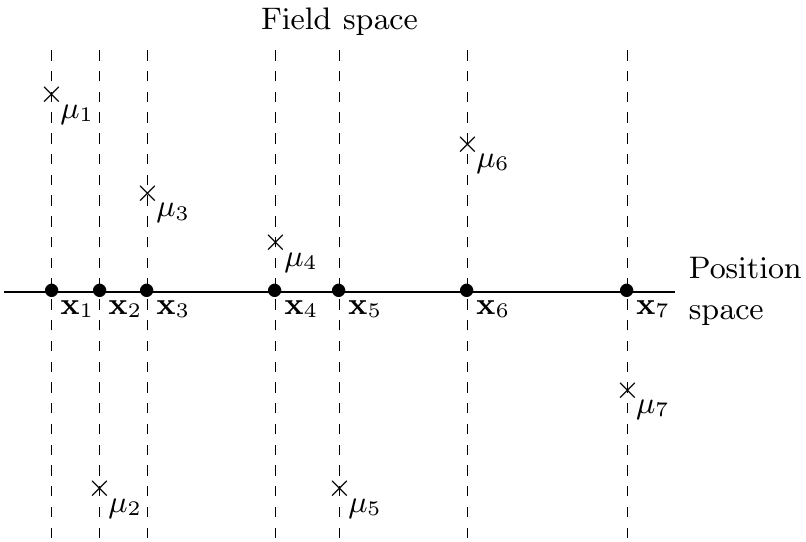}
\caption{ A pictorial depiction of the basis state $|\mu_1,\ldots,\mu_7\rangle$. }
\label{H-picture}
\end{center}
\vspace{-.5cm}
\end{figure}

The generalization of the quantization from quantum mechanics to field theory is
straightforward.  The Poisson bracket of the basic scalar field
variables in Eq. \eqref{basicpb} is  realized as an operator relation on a Hilbert space
with basis states
\be 
| \mu_1,\ldots ,\mu_N\rangle, 
\label{basis}
\ee
where the set of real numbers $\{\mu_j|j=1,\ldots,N\}$ represent scalar field values at the
set of space points $\{\mb{x}_j|j=1,\ldots,N\}$. Figure \ref{H-picture} gives an illustration of a basis state with $N=7$. If two basis states are associated with the same set of points $\{\mb{x}_j|j=1,\ldots,N\}$, the inner product  is given by
\be 
\langle \mu_1',\ldots ,\mu_N'| \mu_1,\ldots ,\mu_N\rangle=
\delta_{\mu_1',\mu_1}\cdot\ldots\cdot \delta_{\mu_N',\mu_N}
\label{ip}
\ee
and it is zero otherwise. 

Equation \eqref{ip} defines an inner product that is background (metric) independent in the same
way  as, for example, that for Ising model spins. A difference is, however, that
for the latter there is a finite dimensional spin space at each
lattice point rather than a real number representing a field value
at each point. In contrast, the usual Fock space quantization uses
the metric dependent Klein-Gordon inner product.  We note that, when
compared with LQG, the set of points on which a
state is based may be called a ``graph" and that there is no scale
associated with the set unless one specifically chooses a uniform lattice.

If we remember that we take $f$ to be a  Gaussian sharply peaked 
at a point $\mb{x}_j$, see Eq. \eqref{gauss}, and if we set
\begin{equation*}
\hat{\Phi}_f(t,\mathbf{x}_j) \equiv \hat{\Phi}_j,  \ \ \ \ \ \ \hat{U}(t,\mathbf{x}_j) \equiv \hat{U}_j, 
\end{equation*}
we have the following actions
\begin{align*}
 \hat{\Phi}_j| \mu_1,\cdots, \mu_n\rangle&= \mu_j|
\mu_1,\ldots ,\mu_n\rangle,\\
 \widehat{U}_j| \mu_1,\ldots \mu_n\rangle &= | \mu_1,\ldots , \mu_j- \lambda/V, \ldots ,\mu_n\rangle. 
\end{align*}
Noting that $f(\mathbf{x}_j-\mathbf{x}')\approx1$ for $(\mathbf{x}_j-\mathbf{x}')^2\ll\sigma^2$ and $f(\mathbf{x}_j-\mathbf{x}')\approx0$ otherwise, it is readily verified that the commutator of these operators leads to a
faithful realization of the corresponding Poisson bracket in Eq. \eqref{basicpb}. Indeed, using $\delta_{j,l}$ for $\delta_{\mb{x}_j,\mb{x}_l}$, we have
\begin{equation}\label{basic-comm}
[ \hat{\Phi}_j, \hat{U}_l] = - \frac{\lambda}{V}\delta_{j,l}\hat{U}_l,	 \ \ \ \ \ \ [ \hat{\Phi}_j, \hat{U}_l ^{\dagger}] =  \frac{\lambda}{V} \delta_{j,l}\hat{U}_l^{\dagger}.
\end{equation}
We note that fixing the parameter $\lambda$  is equivalent to selecting a discreteness scale in field
configuration space, as may be seen from  the action of $\hat{U}$
on basis states given above.

In this representation the momentum operator does not exist because of the ``point" 
nature of the inner product given in Eq. (\ref{ip}) (basis states with different excitations at a
point are orthogonal, as are any states with arbitrary excitations at different points).
 There is, however, an alternative $\lambda$ dependent definition of
a ``momentum" operator given by
\be 
\hat{P}_j  = \frac{1}{2i\lambda}\left(\hat{U}_j - \hat{U}^{\dagger}_j \right),
\label{mom} 
\ee
which can be used to define the kinetic part of the Hamiltonian operator
and any other operators that involve the momentum. For example, the
canonical commutator gets modified from that in usual field theory to
\be
[ \hat{\Phi}_j, \hat{P}_l] = \frac{i}{2V} \delta_{j,l}\left(   \hat{U}_j + \hat{U}^{\dagger}_j    \right).
\label{field-comm}
\ee 
This directly follows from Eqs. \eqref{basic-comm} and \eqref{mom}.

\subsection{Fock operators}
 
We would like to define analogs of the usual field theory creation and 
annihilation operators on the polymer Hilbert space introduced above. 
To do this, we follow as closely as possible the standard constructions, pointing
out along the way where the important differences arise. We begin with the Fourier
decomposition of the classical  field variable and its conjugate momentum
\begin{equation}\label{field-mom}
\begin{split}
\phi(t,\mb{x}) &=\su{\mb{k}}\left[ a_{\mb{k}}f_\mb{k}(t,\mb{x})  + \overline{a}_{\mb{k}}\fbk{}(t,\mb{x})  \right],\\
p(t,\mb{x})&=-i\su{\mb{k}}\ok{}\left[ a_{\mb{k}}f_\mb{k}(t,\mb{x})  - \overline{a}_{\mb{k}}\fbk{}(t,\mb{x})  \right],
\end{split}
\end{equation}
where 
\begin{equation*}
\fk{}(t,\mb{x})= \frac{\mr{e}^{-i\ok{}t+i\mb{k}\mb{x}}}{ \sqrt{2\omega_\mb{k} V} } 
\end{equation*}
are the usual flat space plane waves, $\mb{k}=2\pi\mathbb{Z}^3/L$ with $L^3=V$ and $\ok{}=\sqrt{\mb{k}^2+m^2}$. Given the (Klein-Gordon) inner product  
\begin{equation*}
\langle \chi,\psi\rangle=i\int_{V}\left(\dot{\chi}\overline{\psi}-\chi\dot{\overline{\psi}}\right)\mr{d}^3x
\end{equation*}
for two scalar fields $\chi$ and $\psi$, the plane waves satisfy
\begin{equation*}
	\langle\fk{},\fk{'}\rangle=\langle\fbk{},\fbk{'}\rangle=\delta_{\mathbf{k},\mathbf{k}'},\ \ \ \ \ \ \langle\fk{},\fbk{'}\rangle=0.
\end{equation*}
Here, $\delta_{\mb{k},\mb{k}'}$ denotes the standard Kronecker delta symbol. The expansions in Eq. \eqref{field-mom} give rise to the standard expressions for the creation and annihilation operators
\begin{align*}
\hat{a}_{\mb{k}}&=i\int_V\fbk{}\left(\hat{p}-i\ok{}\hat{\phi}\right)\mr{d}^3x,\\
\hat{a}_{\mb{k}}^{\dag}&=-i\int_V\fk{}\left(\hat{p}+i\ok{}\hat{\phi}\right)\mr{d}^3x.
\end{align*}
We now use this standard procedure to {\it motivate} a definition of the corresponding operators in polymerized QFT using the momentum given in Eq. (\ref{mom});  it is important to note at the outset that 
this may not be the only possible way to proceed and we follow it here simply because it will permit a simple test of the usual field theory limit.  We will also see that, although the $\hat{a}_{\mb{k}}$ are time independent in usual field theory, the corresponding entities in the polymer theory are not, except in the usual field theory limit. 

 To begin, let us consider a subset of points $\{\mb{x}_j|j=1,\ldots,N\}$, which will represent the ``graph" on which  the operators will be defined. (In general, this set can be chosen to be countably infinite, such as a  uniformly spaced lattice but this is not necessary to give the basic prescription.) We define now
\begin{align*}
\Ahk{}&=\frac{iV}{\sqrt{N}} \sum_{j=1}^N \fbk{j}\left(  \hat{P}_j  -i\ok{} \hat{\Phi}_j \right),\\
\Ahdk{}&=-\frac{iV}{\sqrt{N}} \sum_{j=1}^N \fk{j}\left(  \hat{P}_j  +i\ok{} \hat{\Phi}_j \right).
\end{align*}
We note that these operators depend on $\lambda$ (because of $\hat{P}$) and on $t$, and that they are
non-local that is to say they are a sum of operators, one at each point of the graph.  The factor $1/\sqrt{N}$ is necessary to produce the correct  low frequency limit of vacuum expectation values, as we will see below. Also, the explicit time dependence is emphasiced by the non-trivial Heisenberg equations of motion given later.

The polymer Hamiltonian operator  may be defined in the same way as in standard field theory by 
\be
\Hh= \su{\mb{k}}\ok{}\left(\Ahk{}^{\dagger}\Ahk{}+\frac{1}{2}[\Ahk{},\Ahk{}^{\dagger}]\right).
\label{hamilt}
\ee
We will work with this exact form of the Hamiltonian  rather than the normal ordered one
because, as we will see below, it allows us to compute explicitly  a modification to the energy-momentum dispersion relation. In addition, it is useful to keep the full form in any case because
we are after all not working with the usual Fock space quantization where the heuristic argument for
normal ordering is perhaps more compelling. 

Using the  relations (\ref{basic-comm}), it can be seen that the commutator and anti-commutator 
are given by
\begin{align*}
[  \Ahk{}, \Ahdk{}] &= \frac{1}{2N} \sum_{j=1}^N \left(   \hat{U}_j +\hat{U}_j^{\dagger} \right),\\
 \{\hat{A}_{\mathbf{k}}, \hat{A}_{\mathbf{k}}^{\dag}\} &= \frac{V^2}{N}\sum_{j,l=1}^N\overline{f}_{\mathbf{k}j}f_{\mathbf{k}l}   \Big[2 \hat{P}_j\hat{P}_l  +2\omega_{\mathbf{k}}^2\hat{\Phi}_j\hat{\Phi}_l \nn\\
&\quad +i\omega_{\mathbf{k}}  \Big( \{\hat{P}_j,\hat{\Phi}_l  \} -  \{\hat{\Phi}_j,\hat{P}_l  \} \Big)
   \Big],
\end{align*}
where in the first expression we used the definition (\ref{mom}) of the momentum operator. We note here a substantial departure from the usual field theory commutators: the right hand sides  are not constant but  effectively contain an infinite series in powers of the momentum, if the operators $\hat{U}$ and $\hat{U}^{\dag}$ are viewed as arising from the classical Poisson brackets, and an infinite series in powers of the operators $\hat{\Phi}_j$. However, as we will see below, the correct field theory limits can be recovered at low energies. 

\subsection{Vacuum state and expectation values}

Let us now turn to a discussion of an analog of  Fock states in the polymer Hilbert space. Such states will allow us to further probe the commutation rules derived above. The basic idea we follow is that the ``vacuum'' should be a product of oscillator vacuua, one for each point $\mb{x}_j$ of the sample  set of points  $\{\mb{x}_j|j=1,\ldots,N\}$.  Let us first focus on the case of one point, or equivalently the quantum mechanics case, and consider the state
\be 
| G_\mb{k}\rangle_j = \frac{1}{\pi^{1/4}}\su{\mu} \mathrm{e}^{- \ok{}V\mu^2/2}|\mu\rangle_j.
\label{Gj}
\ee
This is just the normalized  ground state of an oscillator of frequency $\ok{}$ at the point $\mb{x}_j$.  
It is normalized to $1$ if we evaluate the sum in Eq. \eqref{Gj} by means of
\begin{equation*}
	\su{\mu}f(\mu)=\sqrt{\ok{}V}\int_{\mathbb{R}}f(\mu)\ \mathrm{d}\mu
\end{equation*}
for an arbitrary integrable function $f$.

Using the state  $| G_\mb{k}\rangle_j  $, we define the  ``polymer Fock vacuum"  on the $N$ point graph
in the mode $\mb{k}$ by 
\begin{equation*}
|0_\mb{k} \rangle = \bigotimes_{j=1}^N | G_\mb{k}\rangle_j.
\end{equation*}
The full vacuum is the tensor product 
\be
|0\rangle = \sideset{}{_\mb{k}}\bigotimes |0_\mb{k} \rangle
\ee

 We will now see that the states $|0_\mb{k}\rangle$ and $|0\rangle$ have the properties expected of the standard quantum field theory  vacuum at sufficiently low frequencies by evaluating several vacuum expectation values. We first define the dimensionless parameter 
\begin{equation*}
\gamma_{\mb{k}} \equiv \ok{}\lambda^2/V
\end{equation*}
to simplify the following expressions. 

The expectation value of the number operator is 
\begin{equation*}
   \langle0_\mb{k}|A_\mb{k}^{\dagger}A_\mb{k}   |0_\mb{k} \rangle =
   \frac{1}{4}-\frac{ \mr{e}^{-\gamma_{\mb{k}} /4}}{2} +\frac{1-\mr{e}^{-\gamma_{\mb{k}}}}{4\gamma_{\mb{k}} }.
\end{equation*}
For the expectation values of the commutator and the anti-commutator we need
\begin{equation*}
    \langle0_\mb{k}| U_j |0_\mb{k} \rangle  = \langle0_\mb{k}| U_j^{\dag} |0_\mb{k} \rangle=\mr{e}^{-\gamma_{\mb{k}}/4 }.
\end{equation*}
This gives
\begin{equation}\label{comm-acomm-expec}
\begin{split}
  \langle0_\mb{k}|[  A_\mb{k}, A_\mb{k}^{\dagger}]  |0_\mb{k} \rangle &=  \mr{e}^{-\gamma_{\mb{k}}/ 4},\\ 
  \langle0_\mb{k}| \{ A_\mb{k}, A_\mb{k}^{\dagger}  \}  |0_\mb{k} \rangle  &=   
  \frac{1}{2} +\frac{1-  e^{-\gamma_{\mb{k}} }}{2\gamma_{\mb{k}} }.
\end{split}
\end{equation}
The expectation value of the Hamiltonian is then
\begin{equation}\label{H-expec}
  \langle0 | \Hh  |0  \rangle = \frac{1}{4}\su{\mb{k}} \ok{} \left( 1+\frac{1-\mr{e}^{-\gamma_{\mb{k}} }}{\gamma_{\mb{k}} }\right).
\end{equation}
Lastly, it is useful to see  the expectation value of the canonical field-momentum commutator 
(\ref{field-comm}), which is 
\begin{equation*}
\lvk [ \hat{\Phi}_j, \hat{P}_l] \rvk = \frac{i}{V}  \delta_{j,l} \ \mathrm{e}^{-\gamma_{\mb{k}} /4}. 
\end{equation*}
This gives a scale dependent modification to the uncertainty relation.
 \begin{figure}
 \vspace{0.8cm}
  \begin{center}
\includegraphics {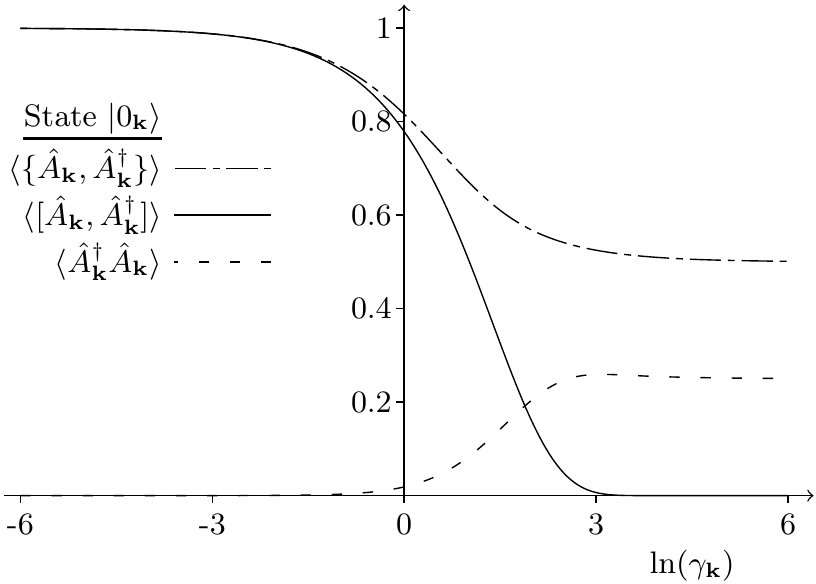}
\caption{The   expectation value of the commutator, the anti-commutator and the number
operator in the state $|0_\mb{k}\rangle$ as a function of $\ln(\gamma_{\mb{k}})$.}
\label{fig0}
\end{center}
\vspace{-.5cm}
\end{figure}
 \begin{figure}
 \vspace{0.8cm}
  \begin{center}
\includegraphics {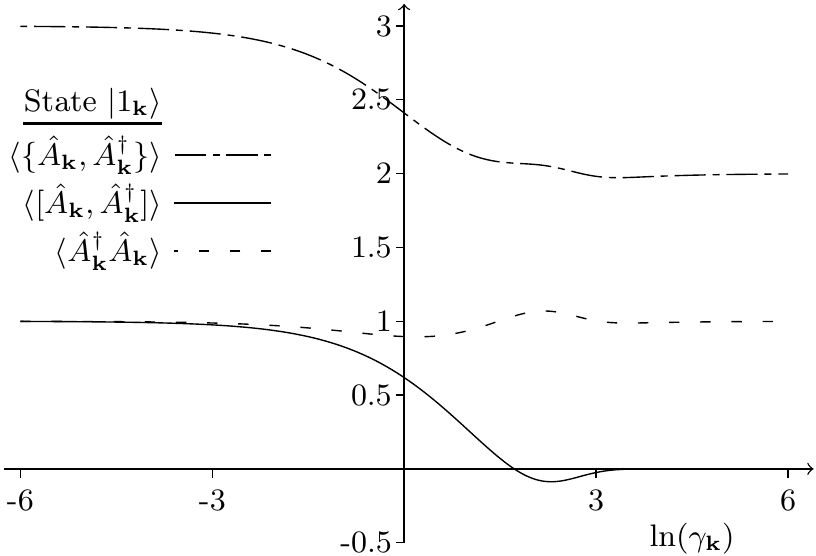}
\caption{The   expectation value of the commutator, the anti-commutator and the number operator in the state $|1_\mb{k}\rangle$ as a function of $\ln(\gamma_{\mb{k}})$.}
\label{fig1}
\end{center}
\vspace{-.5cm}
\end{figure}

It is instructive to consider the low energy limits of these expressions.  For $\gamma_{\mb{k}} \ll 1 $ all the usual field theory vacuum expectation values are recovered.  This justifies our definitions of the vacuum and the creation and annihilation operators. The vacuum becomes an eigenstate of the Hamiltonian in this limit.  

We note that $\lambda$ dependent corrections  set in when $\gamma_{\mb{k}}\sim 1$. 
Figure 2 shows the vacuum expectation value of the commutator, the anti-commutator and the number operator as a function of $\gamma_{\mb{k}}$ over several orders of
 magnitude. From this we see a surprising behavior  at  sufficiently large frequencies: {\it the 
 commutator vanishes and the anti-commutator tends to one half}. This apparent fermionic behavior
 is also clear from Eq. (\ref{comm-acomm-expec}). 

To see whether other states  exhibit this behavior  for large $\gamma_{\mb{k}}$,  it is useful to consider
multiparticle states, which may be defined following the  standard prescription
\begin{equation*}
|n_\mb{k} \rangle =  \frac{1}{\sqrt{\langle n_{\mb{k}}|n_{\mb{k}}\rangle}} \left( \Ahk{}^{\dagger} \right)^n  \rvk. 
\end{equation*}
We computed the same expectation values for up to three particle states using a symbolic algebra
 procedure. The results appear in Figs. 2-5. The main features  evident from each of these graphs
 are that (i) they   indicate the correct low energy limit, and (ii) at high energies the expectation value of the commutator 
 becomes negligible compared to that of the anti-commutator.  We note also that the expectation value of the number operator
 rises with $\gamma_{\mb{k}}$ especially for the two and three particle states.

 \begin{figure}
 \vspace{0.8cm}
  \begin{center}
\includegraphics {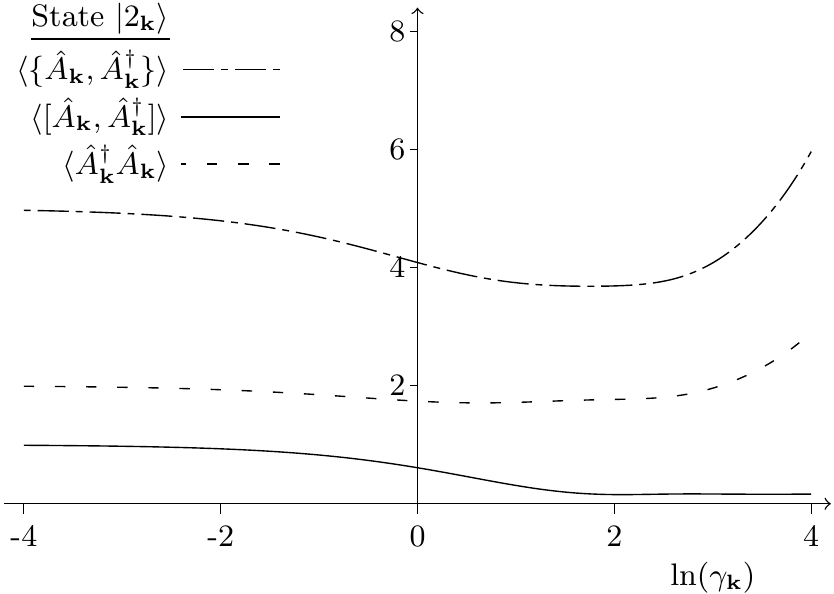}
\caption{The   expectation value of the commutator, the anti-commutator and the number operator in the state $|2_\mb{k}\rangle$ as a function of $\ln(\gamma_{\mb{k}})$.}
\label{fig2}
\end{center}
\vspace{-.5cm}
\end{figure}
 \begin{figure}
 \vspace{0.8cm}
  \begin{center}
\includegraphics {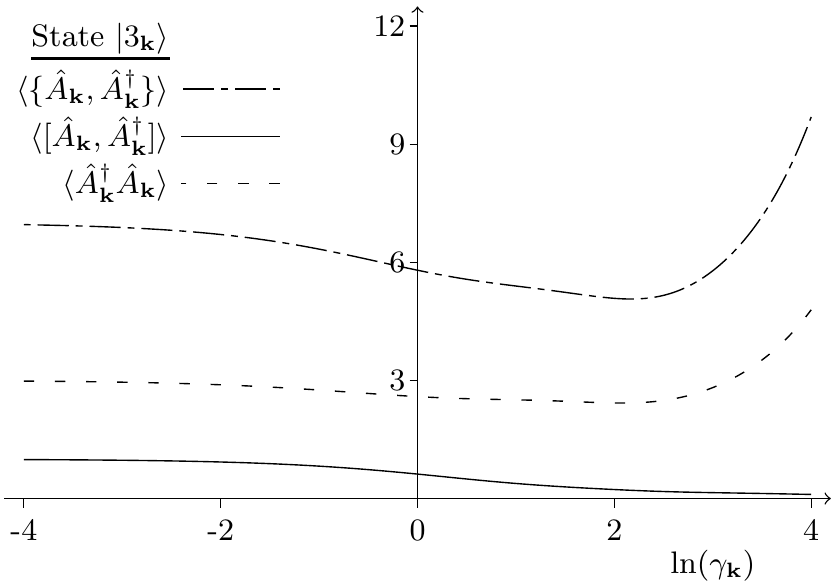}
\caption{The expectation value of the commutator, the anti-commutator and the
number operator in the state $|3_\mb{k}\rangle$ as a function of $\ln(\gamma_{\mb{k}})$.}
\label{fig3}
\end{center}
\vspace{-.5cm}
\end{figure}

\subsection{Time evolution}

 The calculations presented so far have been performed at a single time slice that is to say at a fixed time. Since we have the Heisenberg picture in mind, it is important to
 ask what features of our results are invariant under a time evolution. This is can be checked by 
 calculating commutators with the Hamiltonian  (\ref{hamilt}) such as
 \begin{align*}
[ \hat{H}, [\Ahk{}, \Ahdk{} ] ] &= -i\frac{\ok{}^2\lambda^2}{N}
 \sum_{j=1}^N\left( \fk{j}\Ph_j\Ahk{} + \fbk{j}\Ahdk{}\Ph_j\right),\\
  [\hat{H},\Ahdk{}\Ahk{}]&=i\frac{\ok{}^2\lambda^2}{2N}
 \sum_{j=1}^N \left(\fk{j}\Ph_j\Ahk{}+\fbk{j}\Ahdk{}\Ph_j\right),\\
[ \hat{H},\{\Ahk{},\Ahdk{} \} ]  &= 0.
      \end{align*}
We thus see that at the operator level, only the anti-commutator is a constant of the motion. 
It is also interesting to check the expectation values of the Heisenberg equation of motion. 
For example, for the vacuum  state  $|0_\mb{k}\rangle$ we find 
\begin{align*}
  \lvk[\Hh,\Ahdk{}\Ahk{}]\rvk&=0,\\
    \lvk[\Hh,[\Ahk{},\Ahdk{}]]\rvk&=0.
\end{align*}
It therefore appears that  the fermionic behavior at high energies is time independent, at least for the vacuum state. 
 
\subsection{Modified dispersion relation}

There has been a recent surge in interest in  exploring classical Lorentz violating theories which give rise to modified dispersion relations. This is partly due to the view that a quantum theory of gravity must have this feature at high energies, since such a theory must have a built in scale. Observational bounds have been put  on Lorentz violating terms  in the dispersion relations\cite{ted-rev}, and there have been some derivations of such effects from LQG \cite{GP-disp,Th-Sahl-disp}. It has even been suggested that there is a doubly special relativity \cite{AC-dsr, Lee-dsr} in which modified Lorentz transformations are characterized by not just the speed of light but also by an additional scale.

The quantization we have developed is necessarily Lorentz violating at high energies due
to the scale $\lambda$, so it is interesting to see if we can derive a correction to the dispersion relation.
Using the same quantization method as employed here, this has been done in an effective approach using semiclassical states \cite{HHS-sft}.    However, we  now show, there is an alternative derivation that does not use semiclassical states and that gives a correction
of a lower order.

From the expectation value of the Hamiltonian in Eq. (\ref{H-expec}) we can identify
the $\lambda$ modified
 mode energy
 \be
 E_\mb{k} = \frac{1}{2}\ok{} \left( 1  + \frac{1-
e^{-\gamma_{\mb{k}} } }{\gamma_{\mb{k}} }\right).
 \ee
 Expanding the exponential and remembering that $\omega_\mb{k}^2 = \mb{k}^2 + m^2=|\mb{k}|^2 + m^2$, we find that for $|\mb{k}| \gg m$ 
 \be 
 E_\mb{k}^2 =  | \mb{k}|^2   -
\frac{\lambda^2}{2V} |\mb{k}|^3 + O\left(|\mb{k}|^4\right).
 \ee
It is interesting that this gives a leading order correction proportional to $|\mb{k}|^3$. 
If we take $V$ to be a particle physics mass scale $M$ and   $\lambda$ to be the
Planck scale, the dispersion relation may be written  
\be 
E^2_{\mb{k}} = |\mb{k}|^2 \left[1 - \frac{1}{2} \left( \frac{M}{M_P} \right)^3 \left( \frac{k}{M_P} \right) + \cdots\right], 
\ee
which shows a strong suppression. 

\section{Discussion}
 
 We have developed a background independent ``Fock'' quantization for the scalar field theory by the  relatively straightforward prescription of following standard methods.   Our starting point was a set of 
classical variables  which are essentially the integrated field variable and exponentiated momentum. Utilizing these as the basic variables, we defined  analogs of Fock operators and the vacuum directly in the polymer Hilbert space, and computed   the expectation values of commutators and anti-commutators. The results give the desired low energy limit of usual quantum field theory, and predict a significant departure at  high energy. We also showed that at least for the vacuum  state, the 
observed fermionic  behavior at high energies is time evolution invariant.

 The expectation value of the number operator  in the state $\rvk$ tends asymptotically to $1/4$ as  $\gamma_{\mb{k}}\rightarrow \infty$ and that of the Hamiltonian goes to $0$. Thus, unfortunately, there is no change in the usual vacuum energy problem; we note, however, that we have not diagonalized the Hamiltonian, which is what may be necessary to probe this issue.   Indeed, we  have merely postulated  and justified a choice of vacuum due to its correct low energy limit. This state may not be the true ground state of the Hamiltonian for $\gamma_{\mb{k}} \gg 1$, since we expect that such a state should depend on the parameters   in the Hamiltonian, which in this case is $\lambda$.  An exact  determination of the energy spectrum, however,  requires a   numerical approach due to the $\lambda$ dependence of the Hamiltonian. This  is presently under investigation \cite{HHS-spec}.

 What is the physical interpretation of the apparent fermionic behavior at large $\gamma_{\mb{k}}$? Let us recall that the parameter $\lambda$ is  not a discretization scale associated with space (or time). It is the scale associated with field translations once we have fixed the field momentum operator (\ref{mom}). Thus, the fermionic behavior is not a result of discretizing space (no such behaviour is observed for standard  field theory on a lattice) and it would be puzzling if this were the cause here. Furthermore, there is no indication that the methods we have developed are breaking down at large $\gamma_{\mb{k}}$.   
 
 It is interesting to note that an effective repulsion is  evident when a similar quantization method is applied to the gravitational collapse  of a scalar field with quantum gravity corrections \cite{VH-qgc, KZ-qgc}. In these works a feature of polymer quantization
 is applied to gravitational variables, which effectively lead to a  mass gap at the onset of black hole formation. Although this is not directly an effect of the vanishing of certain commutators at small length scales, it is tempting to speculate that its ultimate origin  may be related. 
  
 In addition to  seeking a better understanding of the high energy behavior,  our results open up several other potentially fruitful directions for further investigation. Among these are  (i) the computation of the propagator, which would require a solution of the Heisenberg equations of motion for the Fock operators or an explicit construction of the exponentiated Hamiltonian, (ii) the free and interacting field theory on a general background, (iii) the quantization of a fermionic  field theory,  (iv) the application to ``weak field'' loop quantum gravity in order to study gravitons, and, finally, (v) applications to Hawking radiation and cosmology.    The latter would be particularly interesting because the evolution of the universe  could itself provide a transition from fermionic to bosonic behavior  during expansion if this method of quantization is the correct one.

 \vskip 2 cm
  
\begin{acknowledgments}
This work was supported in part by the Natural Science and Engineering Research Council of Canada. We thank Jack Gegenberg, Golam Hossain, Gabor Kunstatter and Sanjeev Seahra for discussions.
\end{acknowledgments}

\bibliography{polymer-ft}

\begin{thebibliography}{27}
\expandafter\ifx\csname natexlab\endcsname\relax\def\natexlab#1{#1}\fi
\expandafter\ifx\csname bibnamefont\endcsname\relax
  \def\bibnamefont#1{#1}\fi
\expandafter\ifx\csname bibfnamefont\endcsname\relax
  \def\bibfnamefont#1{#1}\fi
\expandafter\ifx\csname citenamefont\endcsname\relax
  \def\citenamefont#1{#1}\fi
\expandafter\ifx\csname url\endcsname\relax
  \def\url#1{\texttt{#1}}\fi
\expandafter\ifx\csname urlprefix\endcsname\relax\def\urlprefix{URL }\fi
\providecommand{\bibinfo}[2]{#2}
\providecommand{\eprint}[2][]{\url{#2}}

\bibitem[{\citenamefont{Isham}(1995)}]{ISH-strucQG}
\bibinfo{author}{\bibfnamefont{C.~J.} \bibnamefont{Isham}}
  (\bibinfo{year}{1995}), \eprint{gr-qc/9510063}.

\bibitem[{\citenamefont{Ashtekar and Lewandowski}(2004)}]{lqg-AL}
\bibinfo{author}{\bibfnamefont{A.}~\bibnamefont{Ashtekar}} \bibnamefont{and}
  \bibinfo{author}{\bibfnamefont{J.}~\bibnamefont{Lewandowski}},
  \bibinfo{journal}{Class. Quant. Grav.} \textbf{\bibinfo{volume}{21}},
  \bibinfo{pages}{R53} (\bibinfo{year}{2004}), \eprint{gr-qc/0404018}.

\bibitem[{\citenamefont{Thiemann}(2001)}]{lqg-T}
\bibinfo{author}{\bibfnamefont{T.}~\bibnamefont{Thiemann}}
  (\bibinfo{year}{2001}), \eprint{gr-qc/0110034}.

\bibitem[{\citenamefont{Smolin}(2004)}]{lqg-S}
\bibinfo{author}{\bibfnamefont{L.}~\bibnamefont{Smolin}}
  (\bibinfo{year}{2004}), \eprint{hep-th/0408048}.

\bibitem[{\citenamefont{Rovelli}(2008)}]{lqg-R}
\bibinfo{author}{\bibfnamefont{C.}~\bibnamefont{Rovelli}},
  \bibinfo{journal}{Living Rev. Rel.} \textbf{\bibinfo{volume}{11}},
  \bibinfo{pages}{5} (\bibinfo{year}{2008}).

\bibitem[{\citenamefont{Doplicher et~al.}(1995)\citenamefont{Doplicher,
  Fredenhagen, and Roberts}}]{DFR}
\bibinfo{author}{\bibfnamefont{S.}~\bibnamefont{Doplicher}},
  \bibinfo{author}{\bibfnamefont{K.}~\bibnamefont{Fredenhagen}},
  \bibnamefont{and} \bibinfo{author}{\bibfnamefont{J.~E.}
  \bibnamefont{Roberts}}, \bibinfo{journal}{Commun. Math. Phys.}
  \textbf{\bibinfo{volume}{172}}, \bibinfo{pages}{187} (\bibinfo{year}{1995}),
  \eprint{hep-th/0303037}.

\bibitem[{\citenamefont{Husain and Winkler}(2004)}]{HW-cosm}
\bibinfo{author}{\bibfnamefont{V.}~\bibnamefont{Husain}} \bibnamefont{and}
  \bibinfo{author}{\bibfnamefont{O.}~\bibnamefont{Winkler}},
  \bibinfo{journal}{Phys. Rev.} \textbf{\bibinfo{volume}{D69}},
  \bibinfo{pages}{084016} (\bibinfo{year}{2004}), \eprint{gr-qc/0312094}.

\bibitem[{\citenamefont{Husain and Winkler}(2005{\natexlab{a}})}]{HW-qbh1}
\bibinfo{author}{\bibfnamefont{V.}~\bibnamefont{Husain}} \bibnamefont{and}
  \bibinfo{author}{\bibfnamefont{O.}~\bibnamefont{Winkler}},
  \bibinfo{journal}{Class. Quant. Grav.} \textbf{\bibinfo{volume}{22}},
  \bibinfo{pages}{L127} (\bibinfo{year}{2005}{\natexlab{a}}),
  \eprint{gr-qc/0410125}.

\bibitem[{\citenamefont{Husain and Winkler}(2005{\natexlab{b}})}]{HW-qbh2}
\bibinfo{author}{\bibfnamefont{V.}~\bibnamefont{Husain}} \bibnamefont{and}
  \bibinfo{author}{\bibfnamefont{O.}~\bibnamefont{Winkler}},
  \bibinfo{journal}{Class. Quant. Grav.} \textbf{\bibinfo{volume}{22}},
  \bibinfo{pages}{L135} (\bibinfo{year}{2005}{\natexlab{b}}),
  \eprint{gr-qc/0412039}.

\bibitem[{\citenamefont{Hossain
  et~al.}(2009{\natexlab{a}})\citenamefont{Hossain, Husain, and
  Seahra}}]{HHS-cosm}
\bibinfo{author}{\bibfnamefont{G.~M.} \bibnamefont{Hossain}},
  \bibinfo{author}{\bibfnamefont{V.}~\bibnamefont{Husain}}, \bibnamefont{and}
  \bibinfo{author}{\bibfnamefont{S.~S.} \bibnamefont{Seahra}}
  (\bibinfo{year}{2009}{\natexlab{a}}), \eprint{0906.2798}.

\bibitem[{\citenamefont{Hossain
  et~al.}(2009{\natexlab{b}})\citenamefont{Hossain, Husain, and
  Seahra}}]{HHS-sft}
\bibinfo{author}{\bibfnamefont{G.~M.} \bibnamefont{Hossain}},
  \bibinfo{author}{\bibfnamefont{V.}~\bibnamefont{Husain}}, \bibnamefont{and}
  \bibinfo{author}{\bibfnamefont{S.~S.} \bibnamefont{Seahra}},
  \bibinfo{journal}{Phys. Rev.} \textbf{\bibinfo{volume}{D80}},
  \bibinfo{pages}{044018} (\bibinfo{year}{2009}{\natexlab{b}}),
  \eprint{0906.4046}.

\bibitem[{\citenamefont{Ashtekar
  et~al.}(2003{\natexlab{a}})\citenamefont{Ashtekar, Fairhurst, and
  Willis}}]{AFW}
\bibinfo{author}{\bibfnamefont{A.}~\bibnamefont{Ashtekar}},
  \bibinfo{author}{\bibfnamefont{S.}~\bibnamefont{Fairhurst}},
  \bibnamefont{and} \bibinfo{author}{\bibfnamefont{J.~L.}
  \bibnamefont{Willis}}, \bibinfo{journal}{Class. Quant. Grav.}
  \textbf{\bibinfo{volume}{20}}, \bibinfo{pages}{1031}
  (\bibinfo{year}{2003}{\natexlab{a}}), \eprint{gr-qc/0207106}.

\bibitem[{\citenamefont{Halvorson}(2004)}]{halvor}
\bibinfo{author}{\bibfnamefont{H.}~\bibnamefont{Halvorson}},
  \bibinfo{journal}{Stud. Hist. Phil. Mod. Phys.}
  \textbf{\bibinfo{volume}{35}}, \bibinfo{pages}{45} (\bibinfo{year}{2004}),
  \eprint{quant-ph/0110102}.

\bibitem[{\citenamefont{Thiemann}(1998)}]{QSD4}
\bibinfo{author}{\bibfnamefont{T.}~\bibnamefont{Thiemann}},
  \bibinfo{journal}{Class. Quant. Grav.} \textbf{\bibinfo{volume}{15}},
  \bibinfo{pages}{1281} (\bibinfo{year}{1998}), \eprint{gr-qc/9705019}.

\bibitem[{\citenamefont{Varadarajan}(2001)}]{MV-fock}
\bibinfo{author}{\bibfnamefont{M.}~\bibnamefont{Varadarajan}},
  \bibinfo{journal}{Phys. Rev.} \textbf{\bibinfo{volume}{D64}},
  \bibinfo{pages}{104003} (\bibinfo{year}{2001}), \eprint{gr-qc/0104051}.

\bibitem[{\citenamefont{Ashtekar
  et~al.}(2003{\natexlab{b}})\citenamefont{Ashtekar, Lewandowski, and
  Sahlmann}}]{ALS-fock}
\bibinfo{author}{\bibfnamefont{A.}~\bibnamefont{Ashtekar}},
  \bibinfo{author}{\bibfnamefont{J.}~\bibnamefont{Lewandowski}},
  \bibnamefont{and} \bibinfo{author}{\bibfnamefont{H.}~\bibnamefont{Sahlmann}},
  \bibinfo{journal}{Class. Quant. Grav.} \textbf{\bibinfo{volume}{20}},
  \bibinfo{pages}{L11} (\bibinfo{year}{2003}{\natexlab{b}}),
  \eprint{gr-qc/0211012}.

\bibitem[{\citenamefont{Sahlmann}(2007)}]{Sahl-sft}
\bibinfo{author}{\bibfnamefont{H.}~\bibnamefont{Sahlmann}},
  \bibinfo{journal}{Class. Quant. Grav.} \textbf{\bibinfo{volume}{24}},
  \bibinfo{pages}{4601} (\bibinfo{year}{2007}), \eprint{gr-qc/0609032}.

\bibitem[{\citenamefont{Husain et~al.}(2007)\citenamefont{Husain, Louko, and
  Winkler}}]{HLW-hyd}
\bibinfo{author}{\bibfnamefont{V.}~\bibnamefont{Husain}},
  \bibinfo{author}{\bibfnamefont{J.}~\bibnamefont{Louko}}, \bibnamefont{and}
  \bibinfo{author}{\bibfnamefont{O.}~\bibnamefont{Winkler}},
  \bibinfo{journal}{Phys. Rev.} \textbf{\bibinfo{volume}{D76}},
  \bibinfo{pages}{084002} (\bibinfo{year}{2007}), \eprint{0707.0273}.

\bibitem[{\citenamefont{Kunstatter et~al.}(2009)\citenamefont{Kunstatter,
  Louko, and Ziprick}}]{KLZ}
\bibinfo{author}{\bibfnamefont{G.}~\bibnamefont{Kunstatter}},
  \bibinfo{author}{\bibfnamefont{J.}~\bibnamefont{Louko}}, \bibnamefont{and}
  \bibinfo{author}{\bibfnamefont{J.}~\bibnamefont{Ziprick}},
  \bibinfo{journal}{Phys. Rev.} \textbf{\bibinfo{volume}{A79}},
  \bibinfo{pages}{032104} (\bibinfo{year}{2009}), \eprint{0809.5098}.

\bibitem[{\citenamefont{Jacobson et~al.}(2006)\citenamefont{Jacobson, Liberati,
  and Mattingly}}]{ted-rev}
\bibinfo{author}{\bibfnamefont{T.}~\bibnamefont{Jacobson}},
  \bibinfo{author}{\bibfnamefont{S.}~\bibnamefont{Liberati}}, \bibnamefont{and}
  \bibinfo{author}{\bibfnamefont{D.}~\bibnamefont{Mattingly}},
  \bibinfo{journal}{Annals Phys.} \textbf{\bibinfo{volume}{321}},
  \bibinfo{pages}{150} (\bibinfo{year}{2006}), \eprint{astro-ph/0505267}.

\bibitem[{\citenamefont{Gambini and Pullin}(1999)}]{GP-disp}
\bibinfo{author}{\bibfnamefont{R.}~\bibnamefont{Gambini}} \bibnamefont{and}
  \bibinfo{author}{\bibfnamefont{J.}~\bibnamefont{Pullin}},
  \bibinfo{journal}{Phys. Rev.} \textbf{\bibinfo{volume}{D59}},
  \bibinfo{pages}{124021} (\bibinfo{year}{1999}), \eprint{gr-qc/9809038}.

\bibitem[{\citenamefont{Sahlmann and Thiemann}(2006)}]{Th-Sahl-disp}
\bibinfo{author}{\bibfnamefont{H.}~\bibnamefont{Sahlmann}} \bibnamefont{and}
  \bibinfo{author}{\bibfnamefont{T.}~\bibnamefont{Thiemann}},
  \bibinfo{journal}{Class. Quant. Grav.} \textbf{\bibinfo{volume}{23}},
  \bibinfo{pages}{909} (\bibinfo{year}{2006}), \eprint{gr-qc/0207031}.

\bibitem[{\citenamefont{Magueijo and Smolin}(2003)}]{Lee-dsr}
\bibinfo{author}{\bibfnamefont{J.}~\bibnamefont{Magueijo}} \bibnamefont{and}
  \bibinfo{author}{\bibfnamefont{L.}~\bibnamefont{Smolin}},
  \bibinfo{journal}{Phys. Rev.} \textbf{\bibinfo{volume}{D67}},
  \bibinfo{pages}{044017} (\bibinfo{year}{2003}), \eprint{gr-qc/0207085}.

\bibitem[{\citenamefont{Amelino-Camelia}(2002)}]{AC-dsr}
\bibinfo{author}{\bibfnamefont{G.}~\bibnamefont{Amelino-Camelia}},
  \bibinfo{journal}{Int. J. Mod. Phys.} \textbf{\bibinfo{volume}{D11}},
  \bibinfo{pages}{35} (\bibinfo{year}{2002}), \eprint{gr-qc/0012051}.

\bibitem[{\citenamefont{Hossain et~al.}(2010)\citenamefont{Hossain, Husain, and
  Seahra}}]{HHS-spec}
\bibinfo{author}{\bibfnamefont{G.~M.} \bibnamefont{Hossain}},
  \bibinfo{author}{\bibfnamefont{V.}~\bibnamefont{Husain}}, \bibnamefont{and}
  \bibinfo{author}{\bibfnamefont{S.}~\bibnamefont{Seahra}}
  (\bibinfo{year}{2010}), \eprint{to appear}.

\bibitem[{\citenamefont{Husain}(2009)}]{VH-qgc}
\bibinfo{author}{\bibfnamefont{V.}~\bibnamefont{Husain}},
  \bibinfo{journal}{Adv. Science Lett.} \textbf{\bibinfo{volume}{2}},
  \bibinfo{pages}{214} (\bibinfo{year}{2009}), \eprint{0808.0949}.

\bibitem[{\citenamefont{Ziprick and Kunstatter}(2009)}]{KZ-qgc}
\bibinfo{author}{\bibfnamefont{J.}~\bibnamefont{Ziprick}} \bibnamefont{and}
  \bibinfo{author}{\bibfnamefont{G.}~\bibnamefont{Kunstatter}},
  \bibinfo{journal}{Phys. Rev.} \textbf{\bibinfo{volume}{D80}},
  \bibinfo{pages}{024032} (\bibinfo{year}{2009}), \eprint{0902.3224}.

\end{thebibliography}

\end{document}